\documentstyle[aps,floats,prl,twocolumn]{revtex}
\begin{document} 

\title{Phenomenological Understanding of a Transport Regime with Reflection Symmetry in the 
       Quantum Hall System in a Composite Fermion Picture}
\author{Wenjun Zheng, Yue Yu and Zhao-bin Su}
\address{Institute of Theoretical Physics, Chinese Academy of Sciences, Beijing
       100080, P.R.China}
\date{\today}       
\maketitle 

\begin{abstract}
 In this paper, we present a phenomenological picture based on the composite
 fermion theory, in responding to the recent discovery by Shahar et al. of a new transport 
 regime near the transition from a $\nu=1$ quantum Hall liquid to a Hall 
 insulator\cite{Tsui}. In this picture, the seemingly unexpected reflection 
 symmetry in the longitudinal resistivity $\rho_{xx}$ can be understood clearly as due 
 to the symmetry of the gapful excitations which dominate $\sigma_{xx}$ across the transition, 
 and the abrupt change in $\sigma_{xy}$ at the transition. The parameter $\alpha$ in the linear fit of $\nu_0(T)$ 
 in ref\cite{Tsui} is also given a simple physical meaning and the effective 
 mass can be calculated from $\alpha$, which gives
 a reasonable value of several electron band mass. When taking
 into account the result of network model, the almost invariant Hall
 resistivity $\rho_{xy}$ across the transition is also well-understood. 
\end{abstract}

\pacs{ 
\hspace{1cm}
}


In the present research field of quantum Hall effect, there are two hot topics
which are under intensive study both theoretically and experimentally. They are
the anomalous new phase near $\nu= 1/2$ and the transition between a quantum
Hall liquid(QHL) and a Hall insulator(HI)\cite{HI}. The composite fermion(CF) 
picture\cite{Jain}, where the fermionic Chern-Simons field theory\cite{Fradkin}
 is employed, is very successful in describing the former phenomena( and 
 generally other even denominator cases), despite some ambiguities like the 
 divergent effective mass to be clarified. In the CF theory, one can perform 
 a singular gauge transformation by attaching an even number of flux quanta to 
 an electron and obtain the quasiparticle of composite fermion. Near $\nu=1/2$
 the CFs experience a weak effective magnetic field if $\nu$ deviates from 1/2.
 In the seminal paper by Halperin, Lee and Read\cite{HLR}, the above CF 
 theory was also extended to the case with disorder and a qualitatively good 
 phase diagram was suggested. A similar phase diagram was proposed by Kivelson,
 Lee and Zhang\cite{KLZ}, while they derived it from the composite boson(CB) 
 picture, where the bosonic Chern-Simons field theory\cite{SCZ} is used. In 
 this picture, the transition from a QHL to an HI can be intuitively mapped to
  a transition from superfluid to insulator of the CB system. The most 
  attractive idea in this CB picture is the duality transformation between 
  CB and vortex\cite{MPAF}, which offers two bosonic descriptions of the 
  quantum Hall system within the framework of Chern-Simons field theory.
   Considerable attention has been directed to this idea, because of a recently 
  reported experiment by the Princeton group\cite{Tsui}, where substantial 
  evidence for charge-flux duality was discovered. This result implies that 
  the CB picture may be a more promising candidate of the phase transition 
  theory in quantum Hall system, because of its natural inclusion of a charge-flux
  duality\cite{Sondhi}.
 
     In their experiment, Shahar et al.\cite{Tsui} discovered a reflection symmetry 
 in filling factors for the longitudinal resistivity $\rho_{xx}$ between the 1 or $1/3$ QHL and the HI, namely 
\begin{equation}
\label{EQ0}
\rho_{xx}(\nu, T)=1/ \rho_{xx}(\nu_d, T),  
\end{equation}
where $\nu_d=2\nu_c-\nu$ is the dual factor, and $\nu_c=0.562$ is the critical filling factor at the transition for
the situation of $\nu=1$ QHL. The range over which the above equation holds is
 impressively extensive, even beyond the scope of linear transport. There have
 been theoretical attempts to understand this symmetry by means of the 
 self-duality symmetry of the CB theory\cite{Sondhi}. However the microscopic
 justification of this approach is still far from complete.

Further measurement\cite{Tsui} shows (see Fig1 of preprint cond-mat/9706045) 
\begin{equation}
\label{EQ1} 
\rho_{xx}(\nu, T)=\exp[\frac{-\Delta
\nu}{\nu_0(T)}], 
\end{equation} 
where $\rho_{xx}$ is measured in units of its critical value $\rho_{xxc}$,
 $\Delta\nu=\nu-\nu_c$,  
and $\nu_0(T)$ is a $T$-dependent logarithmic slope, which can be fitted to the
 following linear form 
\begin{equation}
\nu_0(T)=\alpha T+\beta, 
\end{equation} 
where $\alpha$ and $\beta$ are sample-dependent parameters. As they pointed 
out, the existence of a non-zero $\beta$ implies the failure of any scaling 
arguments which are extensively used in understanding a continuous phase 
transition \cite{S1}\cite{S2} and the beautiful speculation of super-universality
\cite{SU}. In this sense, they named it a novel transport regime which is 
distinct from both the fully developed QHL and the critical scaling regime.  
This somehow surprising result surely deserves a thorough theoretical 
understanding. However it will not be the question we attempt to deal with 
in this paper.  We will place emphasis on the parameter $\alpha$ which has 
received little attention, and a simple phenomenological picture based on 
the CF theory is proposed, where the surprising relation in eq (\ref{EQ1})
 can be understood easily( excluding the non-zero $\beta$), and the
parameter $\alpha$ is given a clear physical meaning.

    First, let us give some observations which motivate the present study. In
the CF picture of the half-filled Landau level theory, the physics is basically 
symmetric near  $\nu=1/2$. Namely, when $\nu$ deviates from $\nu=1/2$ a little, the CFs
will experience an effective magnetic field  $B^*\propto \Delta \nu$, which 
leads to an energy gap $\Delta=\frac{\hbar eB^*}{m^*}  $, no matter whether
 it is on the left or right side of 1/2. However, this symmetry seems to be broken 
 when a QHL-HI transition occurs near 1/2(we suppose $\nu_c=1/2$ here, and leave the 
 explanation of an experimentally larger $\nu_c$ to the end of this paper).
  For example, in the above experiment, the behavior of $\rho_{xx}$ is quite 
  different at $\nu<\nu_c$ and $\nu>\nu_c$. Since both QHL and HI are gapful
  ( although the nature of a Hall insulator is still elusive to us), it is 
  reasonable to expect something continuous across the critical filling factor 
  $\nu_c$. As pointed out in ref\cite{Tsui}, the Hall resistivity $\rho_{xy}$ 
  is almost constant near the transition. Besides, we emphasize here that the
  longitudinal conductivity $\sigma_{xx}$ will also be continuous and 
  further more show a reflection symmetry near $\nu_c$. The microscopic origin
  of this symmetry is the speculation that the energy gap due to the effective 
  magnetic field $B^*$ in the CF picture is symmetric around 1/2, if we
ignore the possible influence of a $B$-dependent effective mass( we defer a 
discussion of it near the end of this paper). We argue that 
the concept of CF quasi-particle is still useful near the QHL-HI transition where certain amount of
disorder exists. To justify this proposition, we can not approach 1/2 too closely,
so as to maintain an energy gap large enough to avoid being smeared by the disorder.
Under this condition, $\sigma_{xx}$ will be of the form characteristic of a 
thermal excitation hopping mechanism \cite{EXP1}
\begin{equation}
\label{EQ5}
\sigma_{xx}(\nu, T) \propto \exp(-\omega_c^*/k_B T),   
\end{equation}
where $\omega_c^* \propto |\Delta\nu|$, and $\hbar $ is set to unit.
With regard to the above proposition, we further comment that:
This thermal excitation form applied 
to general filling factors is only conditionally reasonable: Generally,
the highest occupied Landau level (LL) for CFs is only partially filled,
and both thermally excited inter LL hoppings and intra LL tunnelings
 contribute to $\sigma_{xx}$. We argue that for $T$ larger than 
the inter LL energy spacing $\omega_c^*$, the former process will be fully 
developed and dominate the conduction over the latter process. While for
T approaching zero, the latter quantum tunneling process should be
taken into account explicitly to explain the observed residue $\sigma_{xx}(T=0)$
and the nonzero parameter $\beta$. In this paper, we confine ourselves
to the situation of $T>\omega_c^*$ and leave the discussion of the 
zero-temperature conduction to a later paper \cite{COMP}.

In spite of the above continuity around 1/2, there should be a sharp change in
the Hall conductivity $\sigma_{xy}$ across the transition point. For $\nu>1/2$
or the QHL phase, we have $\sigma_{xy}=e^2/h~(T\rightarrow 0)$; while for $\nu<1/2$ or
the HI phase, we get $ \sigma_{xy}\rightarrow 0~(T\rightarrow 0)$. This is also consistent with
the well-known "floating up" recipe\cite{FU}, where the QHL-HI transition occurs
at the crossing of the Fermi level with the lowest extended state and 
$\sigma_{xy}$ is determined by the number of extended states below the Fermi level.

    With the above results of conductivity, we can return to the case of 
resistivity, which is directly relevant to the experiments. We can obtain the 
resistivity tensor by conducting an inversion of the conductivity tensor.
 Let us turn to $\rho_{xx}$ first.
\begin{eqnarray}
\rho_{xx}=\frac{\sigma_{xx}}{\sigma_{xx}^2+\sigma_{xy}^2}.
\end{eqnarray}
When $\Delta \nu >0$, 
\begin{eqnarray}
\rho_{xx}&\approx& \frac{\sigma_{xx}}{\sigma_{xy}^2}  \\ \nonumber
         &\propto& \sigma_{xx} \\ \nonumber
         &\propto& \exp(-\omega_c^*/k_B T).
\end{eqnarray}
When $\Delta \nu <0$,
\begin{eqnarray}
\rho_{xx}&\approx& \sigma_{xx}^{-1} \\   \nonumber
         &\propto& \exp(\omega_c^*/k_B T).
\end{eqnarray}
Combining the above results and the relation $\omega_c^*\propto |\Delta\nu|$,
 we can easily identify the reflection symmetry, 
$$ \rho_{xx}(\Delta \nu, T)=\rho_{xx}^{-1}(-\Delta \nu, T).$$
Therefore the reflection symmetry of $\rho_{xx}$ is due to the symmetric energy gap felt by
the CFs and the sharp change in $\sigma_{xy}$ across the transition point. 
Compared with the argument of self-duality symmetry
the present theory can not only explain the symmetric relation in eq (\ref{EQ0}),
 but also give the concrete expression in eq (\ref{EQ1}).
 
Then we turn to $\rho_{xy}$ and see why it remains almost invariant across
the transition.

 When $\Delta\nu >0$,
\begin{eqnarray}
\rho_{xy}&=&\frac{\sigma_{xy}}{\sigma_{xx}^2+\sigma_{xy}^2}   \\  \nonumber
         &\approx& \sigma_{xy}^{-1}  \\  \nonumber
         &=& h/e^2,
\end{eqnarray}
as it should be for a $\nu=1$ QHL.

The situation of $\Delta\nu <0$ is more subtle,
\begin{equation}
\label{EQ2}
\rho_{xy}\approx\frac{\sigma_{xy}}{\sigma_{xx}^2}.
\end{equation}
We note that both the numerator and the denominator of eq (\ref{EQ2}) 
go to zero at the zero temperature limit. In an early calculation\cite{OMEGA}, 
the Kubo formula gave the following result under the condition that $T=0,
 \omega\not =0$,
$$ \sigma_{xx}(\omega)\propto i\omega,$$
while $$ \sigma_{xy}(\omega)\propto \omega^2. $$
So the ratio $\frac{\sigma_{xy}}{\sigma_{xx}^2}$ has a finite value at the 
limit $\omega\rightarrow 0$. However, the more conventional way to compute 
conductivity is to let $\omega$ approach zero first while retain a finite $T$,
 as it is in the present situation. We deem that the ratio will remain a finite value which
depends on the concrete model of HI, such as the network model\cite{NW}. 
In a recent detailed study by Dykhne and Ruzin\cite{DR}, and Ruzin and Feng\cite{RF},
 the finiteness of $\rho_{xy}$ is phrased in terms of a semi-circle law as 
 following
$$ \sigma_{xx}^2+\sigma_{xy}^2=\sigma_{xy}, $$
in the special case of the transition from a $\nu=1$ QHL to an HI. On omitting 
the higher order small term $\sigma_{xy}^2$, we can easily obtain a finite 
ratio at $T\rightarrow 0$
\begin{equation}
\label{EQ4}
\frac{\sigma_{xy}}{\sigma_{xx}^2}\approx 1,
\end{equation}
where the unit of resistivity is $h/e^2$. That is to say, $\rho_{xy}$ is continuous
with the value $h/e^2$ across the transition.

    Now let us compare the above results with the experiments performed by 
Shahar et al.\cite{Tsui}. The value of $\rho_{xy}=h/e^2$ for $\Delta\nu>0$ and 
the finite $\rho_{xy}$ at $\Delta\nu<0$ in eq (\ref{EQ4}) are consistent with the 
experiment, at least in the vicinity of $\nu_c$. This gives support to the present theory. 
However the most important touchstone of the above phenomenological picture 
is the parameter $\alpha$ derived from the measurement of $\rho_{xx}$. If our theory is 
correct, we should be able to obtain an effective mass $m^*$ from the value of 
$\alpha$ by means of the following equation
\begin{equation}
\label{EQ3}
\alpha=\frac{2\pi k_B m^* \nu_c^2}{h^2 n},   
\end{equation}
where $n$ is the density of the two dimensional system. We then use the data 
from ref\cite{Tsui} to estimate the effective mass $m^*$. Substitute 
$n=3\times 10^{10} cm^{-2}$ and $\alpha=0.088 K^{-1}$ into eq (\ref{EQ3}), we get
\begin{eqnarray}
m^*&=&5.36\times 10^{-31}~kg     \\  \nonumber
   &\approx&0.6~m_e        \\   \nonumber
   &\approx&6~m_b, 
\end{eqnarray}
where we have used $m_b\approx 0.1m_e$, and $m_e$ is the mass of a free electron. 
The $m^*$ obtained in this way is consistent\cite{EXP5} with the enhanced effective mass 
 in the CF theory of half-filled Landau level (LL)\cite{HLR} and the relevant 
 experimental results \cite{Mass}. As we believe, this result can serve as a strong support 
 to the present theory, even if a clear understanding of the finite $\beta$ is 
 still absent.

      We then turn to the consideration of the $B$-dependence of $m^*$. In the 
above discussion, we limit ourselves to a small $\Delta\nu$. However the 
experiment shows that the fit in eq (\ref{EQ1}) is good even beyond small 
$\Delta\nu$. In order to understand this in the present theory, we need to 
discuss the concrete $B$-dependence of $m^*$. We notice that $B^*=B-B_{1/2}$ in eq (\ref{EQ5})
 gives a $1/\nu$ factor, which should be cancelled by $m^*$ to ensure a strictly 
 $\nu$-independent $\nu_0(T)$. Therefore we expect the relation $m^*\propto B$ 
 to hold here. This relation can be tentatively explained as a disorder potential induced 
 effective mass which is defined by 
\begin{eqnarray}
\frac{(\Delta k)^2}{2 m^*}&\propto& \frac{(1/{\bf min}[l_B,l_e])^2}{2 m^*}  \\ \nonumber
                          &\propto& \frac{(1/l_B)^2}{2 m^*}  \\   \nonumber
                          &\propto& \bar V_{disorder},
\end{eqnarray}
where $l_B$ is the magnetic length, $l_e$ is the average free path at $B=0$ 
which represents the extent of disorder, $\bar V_{disorder}$ is the evergy 
scale associated with the disorder potential 
which is supposed to be independent of $B$. In arriving at the above results, 
we have used the data from ref\cite{Tsui}: the mobility $\mu=30000~cm^2/V\cdot sec$, 
the density $n=3\times 10^{10}~cm^{-2}$, and calculate 
$$\frac{l_e}{l_B}=\frac{2h\mu n}{e}\approx 2.5~.$$

Therefore
\begin{eqnarray}
m^*&\propto&\frac{1}{l_B^2}  \\   \nonumber
   &\propto& B.
\end{eqnarray}
The above estimation can interpret the extensive fit of eq(\ref{EQ1}) away from
$\nu_c$\cite{EXP4}.

In the end, let us discuss the possible influence of a $\nu_c$ larger than 1/2.
In the situation of the transition from a $\nu=1$ QHL to an HI,
the magnetic field is not very strong, so the mixing between the first and the
second Landau level has to be considered. Therefore, when the first LL is half
filled, a certain number of electrons will fill a small band tail of the second
LL, so the measured filling factor is larger than 1/2. However if we concentrate on the 
first LL and ignore the contribution from the strongly localized electrons in the
second LL, we can still use the formulation for half filled LL and as a result
the use of CF picture is reasonable.


In conclusion, we have proposed a phenomenological picture based on the CF theory,
in order to understand the recent discovery of a new transport regime near the transition
from a $\nu=1$ QHL to an HI. In this picture, the seemingly unexpected reflection 
symmetry can be understood clearly as due to the symmetry of the CF's gapful 
excitations and the sharp change in $\sigma_{xy}$ across the transition. The parameter $\alpha$ which is used to fit 
the experiment is also given a simple physical meaning. Based on the present 
theory the effective mass can be calculated from $\alpha$, which gives a reasonable 
value of several band mass $m_b$. When taking into account the previous 
network model calculations, the almost invariant Hall resistivity $\rho_{xy}$ 
across the transition is also understandable. 

      Before closing, we would like to comment that
the use of a CF picture in the present theory does not exclude a possible explanation
of the experiment based on a CB picture, as it was done in ref\cite{Sondhi}, because
there are certain correspondences between these two pictures. For example, the self-duality
 symmetry in the CB picture corresponds to the particle-hole symmetry in the CF picture.
It is desirable to discuss the relation of the present CF theory to the CB theory which is 
based on self-duality.
It will also be interesting to study what the present theory means to the quantum
phase transition theory for the quantum Hall system, i.e. will it lead to 
a new universal class? These will be the focus of our future work.

 This work is supported in part by the NSF of P. R. China.


\begin{references}
\bibitem{EXP5} We admonish that the quantitative comparison of $m^*$ with the previous
   work in ref[20] is not advisable, because the effects of disorder on $m^*$ are not yet
   well-understood quantitatively.
\bibitem{EXP4} To be exact, the effective mass results from the competition between
               disorder and Coulomb interaction, with the latter's contribution
               being proportional to $\sqrt{B}$, which is relatively small as the experiment
               suggests.
\end{references}
\end{document}